\documentclass{elsart}

\usepackage{natbib}

\usepackage{graphicx}

\usepackage{amssymb}

\journal{New Astronomy Reviews}
\begin{document}

\begin{frontmatter}



\title{A Molecular Gas Study of Low Luminosity Radio Galaxies}


\author[1]{I. Prandoni}, 
\author[2]{R.A. Laing},
\author[1]{P. Parma},
\author[3]{H.R. de Ruiter},
\author[1]{F.M. Montenegro-Montes} and
\author[2]{T.L. Wilson}

\address[1]{INAF - Istituto di Radioastronomia, Bologna, Italy} 
\address[2]{European Southern Observatory, Garching b. Munch\"en, Germany} 
\address[3]{INAF - Osservatorio Astronomico di Bologna, Bologna, Italy} 

\begin{abstract}
We discuss CO spectral line data of a volume-limited sample of 23 nearby 
($z<0.03$) low luminosity radio galaxies, selected from the B2 catalogue. 
We investigate whether the CO properties of our sample are correlated with 
the properties of the host galaxy, and in particular with the dust component. 
We find strong evidences for a physical link between the dust disks probed by 
HST in the galaxy cores and the molecular gas probed by the CO spectral 
lines, which in two cases display a double-horn shape, consistent with 
ordered rotation. 
On the other hand, from a preliminary comparison with other samples of radio 
sources we find no significant differences in molecular gas properties between 
FRI and FRII radio sources.
In order to confirm the suggestion that the CO is dynamically associated with 
the core dust disks, the most suitable sources of our sample will be proposed 
for interferometric imaging at PdBI.
\end{abstract}


\end{frontmatter}

\section{Background}\label{sec-back}
The fuelling of relativistic jets in radio-loud active galactic nuclei is still
not fully understood.  It is clear that energy is released in the vicinity of a
supermassive black hole, but whether the mechanism is direct electromagnetic
extraction of rotational kinetic energy (e.g. \citealt{BZ77}; 
\citealt{Ketal02}) or more closely related to the process of accretion 
(e.g. \citealt{BP82}; \citealt{Hetal03}) remains a matter of 
debate. The ratio of energy
flux in jets to that radiated by an accretion disk varies by orders of 
magnitude between different classes of AGN. Here, we are concerned 
with low-luminosity (FRI) radio galaxies (\citealt{FR74}). 
These are known to contain very massive black holes, but show very little 
evidence for emission from accretion disks, their nuclear luminosities 
being a small fraction of that expected from accretion at the Eddington rate, 
$L/L_{\rm Edd} < 10^{-4}$ (\citealt{Metal04}). 
They form the parent population for nearby BL Lac objects and must
therefore produce highly relativistic jets on sub-pc scales. There is direct
evidence for relativistic motion on parsec scales in FR\,I jets 
(\citealt{Getal01}) 
and for smooth deceleration from relativistic to sub-relativistic
speeds on scales of 1 -- 10\,kpc (\citealt{Letal99}).
FR\,I radio sources are located in fairly normal elliptical galaxies, 
invariably containing hot, X-ray emitting plasma 
(thought to confine the jets on large scales), but little ionized, 
line-emitting material at $\approx10^4$\,K. It has recently become clear that 
they may also contain substantial amounts of cool gas and dust 
(e.g. \citealt{deRetal02}; \citealt{VZ05}; \citealt{Letal00}; 
\citealt{Letal03}).  
Dust is observed in 53\% of the B2 sample of nearby radio galaxies 
(mostly FRI; see below) and the dust mass is correlated with radio power 
(\citealt{deRetal02}). There is also a connection between the dust-lane 
morphology (disk/irregular) and the presence of jets, and some tendency for 
dust lanes and jets to be orthogonal (\citealt{deRetal02}; 
\citealt{VZ05}).
These associations argue that accretion of cool gas may indeed power the
radio jets.
The next step is to understand the dynamics of the cool gas. \citet{Letal00}
detected $^{12}$CO (1 $\rightarrow$ 0) and (2 $\rightarrow$ 1) emission from 
the FR\,I radio galaxies 3C\,31 and 264 with the IRAM 30m Telescope and 
established that the line profiles indicate disk rotation. 
Interferometric observations of 3C\,31 by \citet{Oetal05} showed
that the CO coincides spatially with the dust disk observed by HST 
(\citealt{Metal99}) and is in ordered rotation.  
These authors suggest that the cool gas
is in stable orbits. \\
In order to increase the number of FR\,I radio galaxies with CO 
observations and so to improve our knowledge of their molecular gas properties,
we are studying a volume-limited sample of 23 nearby 
($z<0.03$) low luminosity radio galaxies, selected from the B2 
catalogue (\citealt{Cetal75}). We notice that for 16 of such objects HST 
imaging is available (\citealt{Cetal00}).
The CO properties of this sample are compared to the 23 $z<0.03$ 
3C radio galaxies studied by 
\citet{Letal03} and to the 18 $z<0.0233$ (or $v<7000$ km/s) 
UGC galaxies with radio jets, studied by \citet{Leonetal03}. 
We notice that the three samples are partially overlapping.

\begin{table}[t]
\caption{IRAM 30m CO line measurements
\label{tab-meas}}
\footnotesize
\vspace{0.3truecm}
\begin{tabular}{crcrrcrr}
\hline
\multicolumn{1}{c}{Source} & \multicolumn{3}{c}{$^{12}$CO(1$\rightarrow$0)} & 
\multicolumn{3}{c}{$^{12}$CO(2$\rightarrow$1)} 
& \multicolumn{1}{c}{$\log{M_{H_2}}$}\\
\multicolumn{1}{c}{} & \multicolumn{1}{c}{$\Sigma T_a^* dv$}  & 
\multicolumn{1}{c}{$\Delta v_{\tt FWHM}$} & 
\multicolumn{1}{c}{$\frac{T_{\rm peak}}{T_{\rm rms}}$} & 
\multicolumn{1}{c}{$\Sigma T_a^* dv$}  & 
\multicolumn{1}{c}{$\Delta v_{\tt FWHM}$} & 
\multicolumn{1}{c}{$\frac{T_{\rm peak}}{T_{\rm rms}}$} & 
\multicolumn{1}{c}{} \\
\multicolumn{1}{c}{} & \multicolumn{1}{c}{K km/s}  & 
\multicolumn{1}{c}{km/s} & 
\multicolumn{1}{c}{} & 
\multicolumn{1}{c}{K km/s}  & 
\multicolumn{1}{c}{km/s} & 
\multicolumn{1}{c}{} & 
\multicolumn{1}{c}{M$\odot$} \\
\hline
$0120+33$ & $<0.35$ & - & - & $<0.61$ & - & - & $<7.7$ \\
$0149+35$ & $1.34$ & 545 & 5.4 & $3.15$ & 551 & 7.7 & $8.3$ \\
$0258+35$ & $0.88$ & 382 & 6.6 & $0.21$ & 127 & 2.4 & $8.1$ \\
$0326+39$ & $<0.35$ & - & - & $0.64$ & 171 & 1.9 & $<8.0$ \\
$0331+39$ & $1.35$ & 807 & 3.6 & $0.90$ & 679 & 3.4 & $8.5$ \\
$1122+39$ & $6.84$ & 670 & 23.0 & $5.93$ & 672 & 16.7 & $8.2$ \\
$1217+29$ & $0.58$ & 346 & 3.3 & $<0.87$ & 335 & 3.1 & $6.1$ \\
$1321+31$ & $<0.35$ & - & - & $1.24$ & 297 & 2.7 & $<7.7$ \\
$1615+35$ & $<0.35$ & - & - & $0.52$ & 129 & 4.1 & $<8.2$ \\
\hline
\multicolumn{8}{l}{$M_{H_2}$ derived from CO(1$\rightarrow$0) 
measures and assuming $H_0=100$ km/s/Mpc}
\end{tabular}
\vspace{0.5cm}
\end{table}

\section{Line Observations and Measurements} \label{obs}
We used the IRAM 30m telescope to search for emission in the 
(1$\rightarrow$0) and (2$\rightarrow$1) transitions of $^{12}$CO in 9
B2 radio galaxies of the $z<0.03$ volume-limited sample, described above. 
We used receivers A100 and B100 connected to the 1 MHz filter bank in $2\times
512$ MHz blocks together with receivers A230 and B230 with the 4 MHz filter 
bank in $2\times 1$ GHz blocks. 
After averaging the outputs of the two pairs of receivers
we got noise levels of $T_{\rm rms}\sim 0.5-1$ 
mK and $\sim 1-2$ mK 
respectively ($1\sigma$, $\Delta v \sim 40$ km/s).
The data were reduced with the CLASS package and line fluxes were
measured by numerically integrating over the channels in the line profile.
Line widths were measured as full widths at 50\% of the peak. 
A source was considered detected when both (1$\rightarrow$0) and 
(2$\rightarrow$1) emission lines have $T_{\rm peak}>2 T_{\rm rms}$, 
with at least one having $T_{\rm peak}>3T_{\rm rms}$. 
In case of non detections, upper limits were 
calculated (see \citealt{Eetal05}). A summary of our CO line measurements 
is given in Table~\ref{tab-meas}.
Line widths of the 5 detected sources are of the order of 500 km/s, 
and in a few 
cases lines show a double-horn structure, indicating rotating CO disks.
$H_2$ molecular masses were derived as in \citet{Letal00}.

\begin{figure}[t]
\resizebox{8.5cm}{!}{\includegraphics{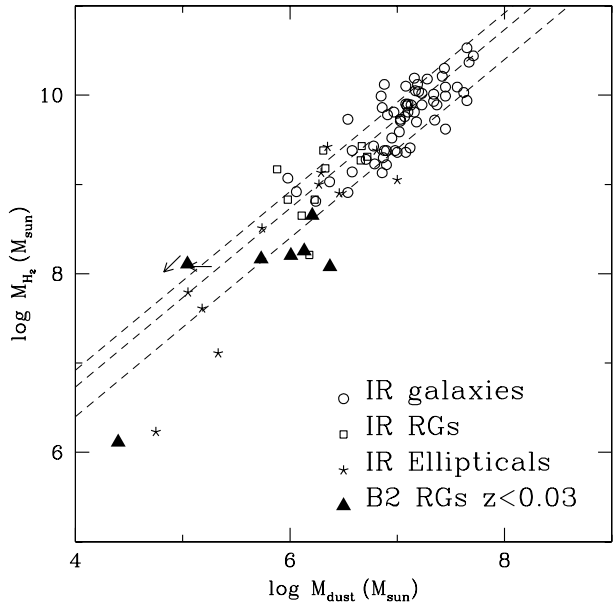}}
\hfill
\parbox[b]{45mm}{
\caption{Dust mass (derived from IRAS observations) vs. $H_2$ mass for 
several IR-selected samples: 
IRAS luminous galaxies (empty circles); IRAS radio galaxies 
(empty squares); IRAS ellipticals (asterisks). Filled triangles and arrows
refer to IRAS $z<0.03$ B2 radio galaxies. Dashed lines indicate the 
$M_{\rm H_2}/M_{\rm dust}=540 \pm 290$ 
relation found by \citet{Setal91}.  
\label{fig-H2dust}}}
\vspace{0.5cm}
\end{figure}

\section{Molecular gas and dust properties of the B2 $z<0.03$ 
sample}\label{sec-gasprop}

In this section we want to investigate whether the CO properties of our sample
are correlated with the properties of the host galaxy, 
and in particular with the dust component.
We notice that to our measurements (see Table~\ref{tab-meas}) we have 
added CO line measurements for other 7 galaxies, observed as part of the
UGC and 3C samples (\citealt{Leonetal03}; \citealt{Letal03}, see 
Sect~\ref{sec-back}). This means that our analysis can rely on 16 of the 23 
galaxies (i.e. $\sim 70\%$ of the whole sample). \\
In Fig.~\ref{fig-H2dust} we show the relation between 
the molecular mass content, $M_{H_2}$, and the content in ``warm'' dust 
(derived from far-infrared 60 and 100 $\mu$m observations), 
$M_{\rm dust}$, found for several 
IR-selected galaxy samples: luminous IRAS galaxies (empty circles; 
\citealt{Setal91}); IRAS radio galaxies (empty squares; \citealt{Metal93}; 
\citealt{Eetal05}); IRAS ellipticals (asterisks; \citealt{Wetal95};
\citealt{Y02}; \citealt{Y05}). 
Also indicated is the $M_{\rm H_2}/M_{\rm dust}=540 \pm 290$ relation 
measured for 
IRAS luminous galaxies (dashed lines, \citealt{Setal91}).
In our sample of 16 B2 $z<0.03$ radio galaxies with CO line measurements, 
we have 9 objects detected by IRAS. 
Such objects are indicated in Fig.~\ref{fig-H2dust} 
by filled triangles and/or arrows (in case of upper limits).
We notice that IRAS-detected sources in our
sample are typically also detected in CO (only 1 CO upper limit is present in
the figure\footnote{upper 
limits in $M_{\rm dust}$ are defined for objects detected by IRAS 
only at 60 $\mu$m.}), indicating an association between dust and 
molecular gas components.\\
Our sample can be used to 
probe the $M_{\rm H_2}-M_{\rm dust}$ relation at lower masses than typical for 
luminous IR galaxies and other IR-selected radio galaxies. Our 
low luminosity radio galaxies show a large spread in dust masses and are not 
well represented by the relation holding at larger masses. 
In particular, several galaxies show larger dust masses than expected. 
While this result can suggest a departure from the relation at very low mass 
ranges (see also 
IRAS ellipticals, asterisks), it can also possibly be due to the different 
scales probed by IRAS (the whole galaxy) and CO observations (inner 5-10 kpc 
at the $^{12}$CO(1$\rightarrow$0) observing frequency). 

\begin{figure}[t]
\centering
\begin{minipage}[c]{0.5\textwidth}
\centering
\includegraphics[width=5.5cm,angle=-90]{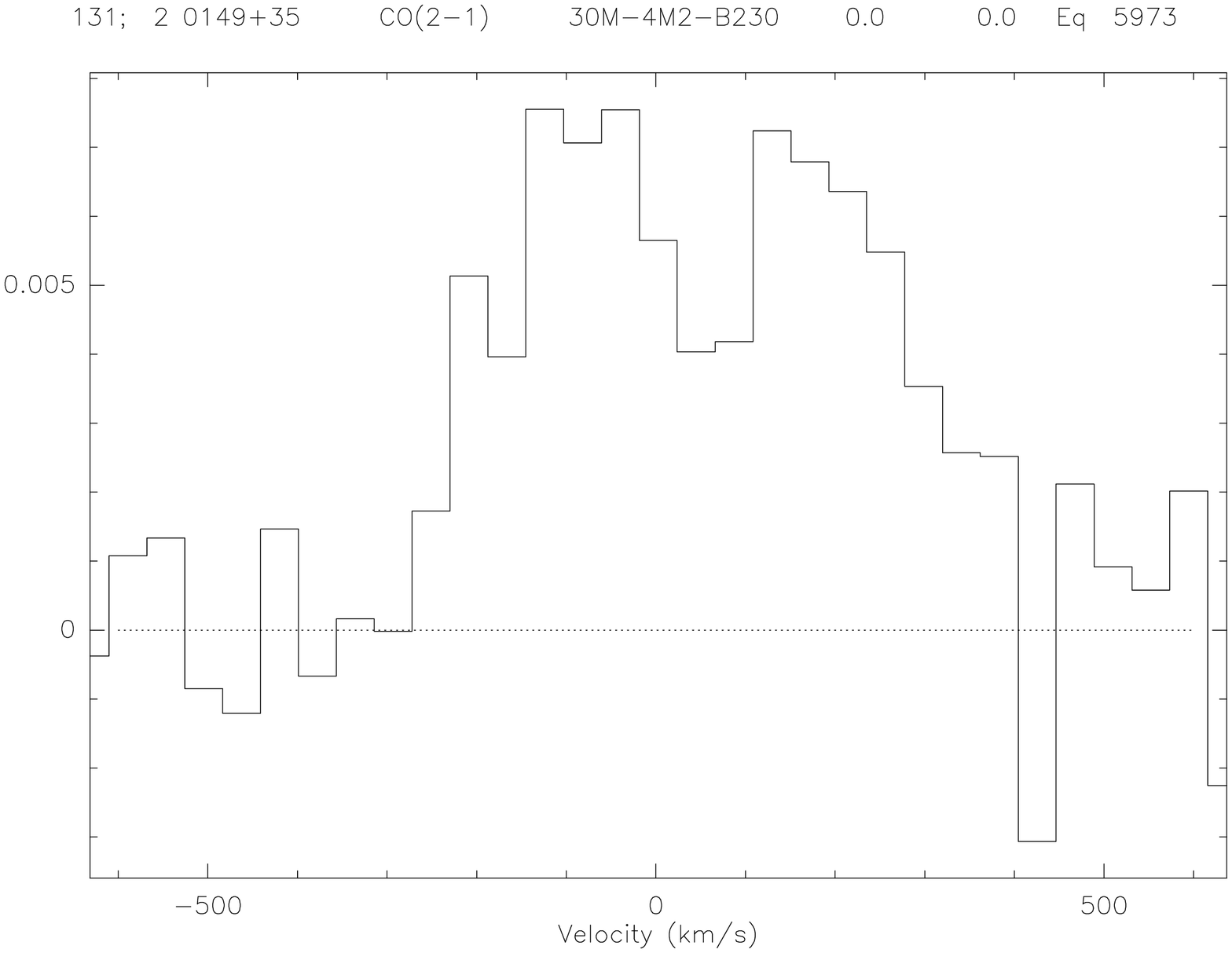}
\end{minipage}%
\begin{minipage}[c]{0.5\textwidth}
\centering
\includegraphics[width=5.5cm]{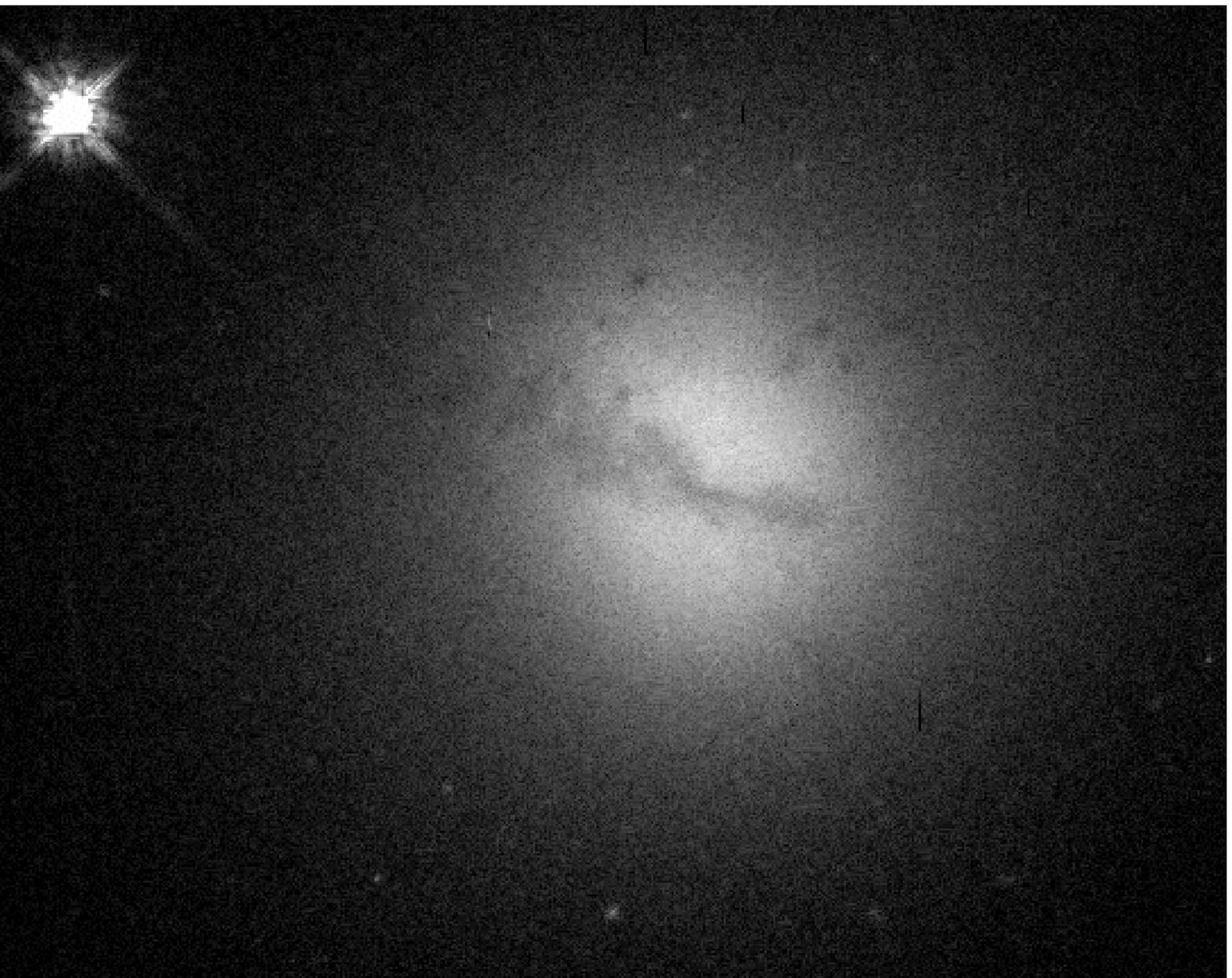}
\end{minipage}
\caption[]{{\it Left:} $^{12}$CO(2$\rightarrow$1) spectrum obtained at IRAM 
30m for the B2 radio galaxy 0149+35, showing a double-horn line. {\it Right:} 
HST image of 0149+35, showing a dusty rotating disk. 
\label{fig-coline}}
\end{figure}

\begin{figure}[t]
\resizebox{8.5cm}{!}{\includegraphics{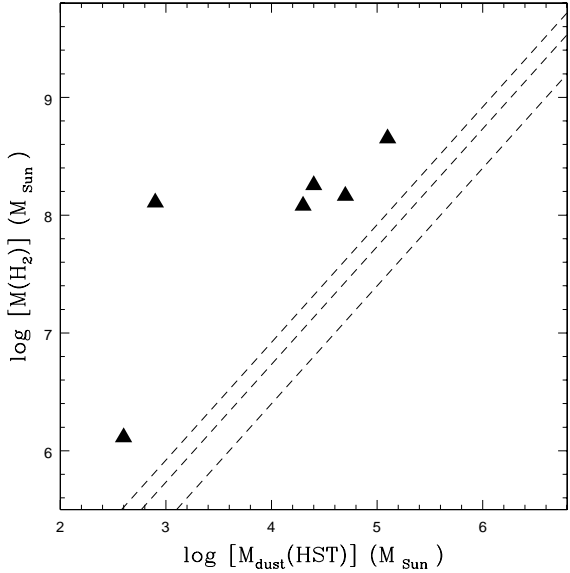}}
\hfill
\parbox[b]{45mm}{
\caption{Dust mass (derived from HST observations) vs. $H_2$ mass for the 
$z<0.03$ B2 radio galaxies with HST dust and CO line detections. 
The $M_{\rm H_2}/M_{\rm dust}=540\pm 290$ relation 
found for IRAS luminous galaxies (\citealt{Setal91}) is overplotted as 
reference (dashed lines). 
\label{fig-H2dust2}}}
\vspace{0.5cm}
\end{figure}

More meaningful it could be from this point of view a comparison between 
CO and HST high resolution
observations, which are approximately sensitive to the same scale (the 
galaxy core). \\
HST observations are available for 12 of the 16 galaxies with CO line 
measurements in our sample. We find strong evidences for a physical link 
between the dust component probed by HST and the molecular gas 
probed by CO: CO is detected only in those galaxies 
showing dust in HST images and double-horn CO lines are found in two 
galaxies, both showing HST rotating dusty disks (see example in 
Fig.~\ref{fig-coline}). Such evidences reinforce previous indications 
by \citet{Letal00} and \citet{Oetal05}. \\
In order to check whether any quantitative relation can be derived between 
molecular gas and optical dust masses in our sample, 
we plot in Fig.~\ref{fig-H2dust2} $M_{H_2}$ against the optical dust mass, 
as derived from HST observations 
($M_{\rm dust}(HST)$, see \citealt{deRetal02}).
It is clear that HST-derived dust masses are a factor 10-100 lower than
masses derived from IRAS observations, probably reflecting the 
different galaxy scales probed by HST and IRAS. Typically we have 
$M_{\rm H_2}/M_{\rm dust}(HST) \sim 3000 \div 6000$, except for a source
showing an anomalously large molecular-gas-to-dust ratio 
(of the order of $10^5$). \\ 
However only 6 objects have reliable CO measures (we discard upper 
limits) and a reliable statistical analysis must await further CO line
single-dish observations. In order to probe the gas dynamics and 
confirm the existence 
of a dynamical link between CO and HST dusty disks, the most 
suitable B2 radio sources will also be proposed for interferometry at 
Plateau de Bure. 

\section{Comparison with other radio samples}\label{sec-comp}
\begin{figure}[t]
\resizebox{8.5cm}{!}{\includegraphics{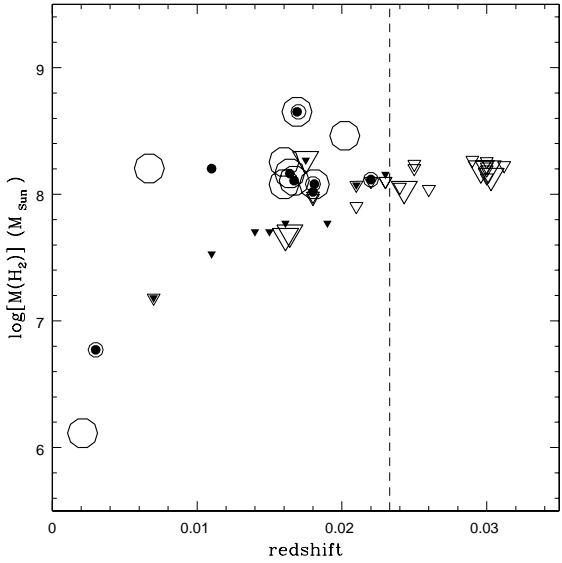}}
\hfill
\parbox[b]{45mm}{
\caption{$H_2$ mass vs redshift for the 
$z<0.03$ B2 radio galaxies (large empty symbols); 
for $z<0.03$ 3C radio galaxies (small empty symbols); and for $z<0.0233$ UGC 
galaxies (filled symbols). 
Circles refer to detections and triangles to upper limits. 
The dashed line at $z=0.0233$ indicates the redshift limit to 
which the three samples overlap.
\label{fig-H2massz}}}
\vspace{0.5cm}
\end{figure}

We have finally 
compared the B2 $z<0.03$ radio galaxy sample to the $z<0.03$ 3C 
and the $z<0.0233$ UGC radio samples (\citealt{Letal03}; \citealt{Leonetal03}).
The aim is to check whether the molecular gas properties 
are different for different types of radio sources. In particular we want to
confirm the findings of \citet{Eetal05}, who claim that FRII galaxies
are characterized by lower CO detection rates than FRI and compact radio 
sources.  \\
To take into account the differences in volume, 
we limited the comparison to all objects with $z \leq 0.0233$. In such range
we find no significant difference (see Fig.~\ref{fig-H2massz}): molecular 
gas masses are very similar, approximately spanning the 
range $10^7 - 10^9$ M$_\odot$, with upper limits varying 
from $\sim 10^7$ M$_\odot$
to $\sim 10^8$ M$_\odot$, depending on distance. 
Also CO detection rates 
do not differ significantly in the three samples: $47\pm 17\%$, 
$33\pm 17 \%$, and $75\pm 25 \%$,  for the UGC, 3C and B2 samples respectively.
We notice however that the three samples partially overlap and all the 
3C FRII sources studied by Lim et al. (2003) are at $z>0.0233$. 
While this can explain
the results of Evans et al., as a consequence of a distance bias against 
FRII radio galaxies, it is clear that deeper CO observations are needed
to better constrain the molecular gas properties of FRI and FRII radio sources.

\end{document}